\documentclass[aps,pra,superscriptaddress,twocolumn]{revtex4-2}

\usepackage{graphicx}
\usepackage{amsmath}
\usepackage{amssymb}
\usepackage{hyperref}
\usepackage[utf8]{inputenc}
\usepackage{mathtools}
\usepackage[english]{babel}
\usepackage{bbm}
\hypersetup{colorlinks=true, linkcolor=blue, citecolor=blue, urlcolor=blue}
\usepackage{xcolor}
\usepackage{braket}
\usepackage[normalem]{ulem}
\usepackage[utf8]{inputenc}
\usepackage{float}
\usepackage[export]{adjustbox}
\usepackage{subcaption}
\usepackage{caption}
\usepackage[justification=raggedright,singlelinecheck=false]{caption}

\usepackage{xcolor}
\newcommand{\new}[1]{\textcolor{black}{#1}}

\begin{document}
\title{Fourier Neural Operators for Learning Dynamics in Quantum Spin Systems}

\author{Freya Shah}
\affiliation{Department of Computing + Mathematical Sciences (CMS), California Institute of Technology (Caltech), Pasadena, CA 91125, USA}
\affiliation{Ahmedabad University, Ahmedabad 380015, Gujarat, India}

\author{Taylor L. Patti}
\affiliation{NVIDIA, Santa Clara, CA 95051, USA}

\author{Julius Berner}
\affiliation{Department of Computing + Mathematical Sciences (CMS), California Institute of Technology (Caltech), Pasadena, CA 91125, USA}

\author{Bahareh Tolooshams}
\affiliation{Department of Computing + Mathematical Sciences (CMS), California Institute of Technology (Caltech), Pasadena, CA 91125, USA}

\author{Jean Kossaifi}
\affiliation{NVIDIA, Santa Clara, CA 95051, USA}

\author{Anima Anandkumar}
\affiliation{Department of Computing + Mathematical Sciences (CMS), California Institute of Technology (Caltech), Pasadena, CA 91125, USA}

\begin{abstract}

Fourier Neural Operators (FNOs) excel on tasks using functional data, such as those originating from partial differential equations. Such characteristics render them an effective approach for simulating the time evolution of quantum wavefunctions, which is a computationally challenging, yet coveted task for understanding quantum systems. In this manuscript, we use FNOs to model the evolution of random quantum spin systems, so chosen due to their representative quantum dynamics and minimal symmetry. We explore two distinct FNO architectures and examine their performance for learning and predicting time evolution using both random and low-energy input states. \new{We find that standard neural networks in fixed dimensions, such as U-Net, exhibit limited ability to extrapolate beyond the training time interval, whereas FNOs reliably capture the underlying time-evolution operator and generalize effectively to unseen times.} Additionally, we apply FNOs to a compact set of Hamiltonian observables ($\sim\text{poly}(n)$) instead of the entire $2^n$ quantum wavefunction, which greatly reduces the size of our inputs and outputs and, consequently, the requisite dimensions of the resulting FNOs. Moreover, this Hamiltonian observable-based method demonstrates that FNOs can effectively distill information from high-dimensional spaces into lower-dimensional spaces. \new{Using this approach, we perform numerical experiments on a 20-qubit system and extrapolate Hamiltonian observables to twice the training time with a relative error of $5.8\%$. Notably, relative to numerical time-evolution methods, FNO achieves an inference speedup of approximately $10^{4}\times$ for 20-qubit systems, underscoring its computational efficiency at larger system sizes.} The extrapolation of Hamiltonian observables to times later than those used in training is of particular interest, as this stands to fundamentally increase the simulatability of quantum systems past both the coherence times of contemporary quantum architectures and the circuit-depths of tractable tensor networks. 

\end{abstract}

\maketitle

Simulating the dynamics of quantum systems has been a long-standing goal for the scientific community, underpinning Feynman's initial proposition of quantum computing~\cite{feynman2018simulating,zalka1998efficient,lloyd1996universal}. Learning and predicting the behavior of intricate quantum spin systems presents a significant challenge due to their inherent superpolynomial time complexity \cite{georgescu2014quantum}. Controllable quantum systems, such as quantum simulators or other quantum computers, represent a promising pathway for simulating complex quantum systems, as they share similar dynamics and large Hilbert spaces \cite{daley2022practical,altman2021quantum,ebadi2021quantum,tacchino2020quantum,trivedi2024quantum,king2024computational}. However, current quantum computing technologies face significant limitations \cite{schlosshauer2019quantum}. In the present Noisy Intermediate-Scale Quantum (NISQ) era, quantum computers are restricted to a limited number of qubits and substantial error rates due to decoherence and operational imperfections \cite{preskill2018quantum,fellous2021limitations}. These coherence and scalability issues constrain the capacity of quantum computers to simulate large and complicated spin systems, particularly over long timescales. As a result, achieving substantial and accurate results in the simulation of these systems remains elusive.

Advances in quantum modeling have introduced promising new approaches for quantum simulation, addressing limitations of traditional techniques. For instance, tensor methods like the Density Matrix Renormalization Group (DMRG) are capable of simulating larger quantum systems \cite{white1992density,schollwock2005density,orus2019tensor,panagakis2024tensor}. Tensor methods are highly effective for certain applications, such as studying ground state properties in one-dimensional systems but they encounter significant limitations when extended to systems with higher dimensions or greater levels of entanglement. Likewise, machine learning techniques based on neural networks, such as Neural-Network Quantum States (NQS) and Heisenberg Neural Networks (HENN), capture the essence of large quantum systems with a smaller-dimensional model. NQS provides a compact representation of many-body quantum states with artificial neural networks, capturing intrinsically nonlocal correlations and improving scalability over conventional approaches \cite{carrasquilla2017machine,carleo2017solving,torlai2018neural,lange2024architectures}, while HENN reconstructs time-dependent quantum Hamiltonians from local measurements, employing a physics-informed loss function based on the Heisenberg equation of motion and achieving high tomographic fidelity with sparse data \cite{han2021tomography,mohseni2022deep}. However, such machine learning-based approaches have marked limitations in accuracy, particularly when they model large quantum systems and long evolution times.

\begin{figure*}[htb]
\centering
\includegraphics[width=0.8\textwidth]{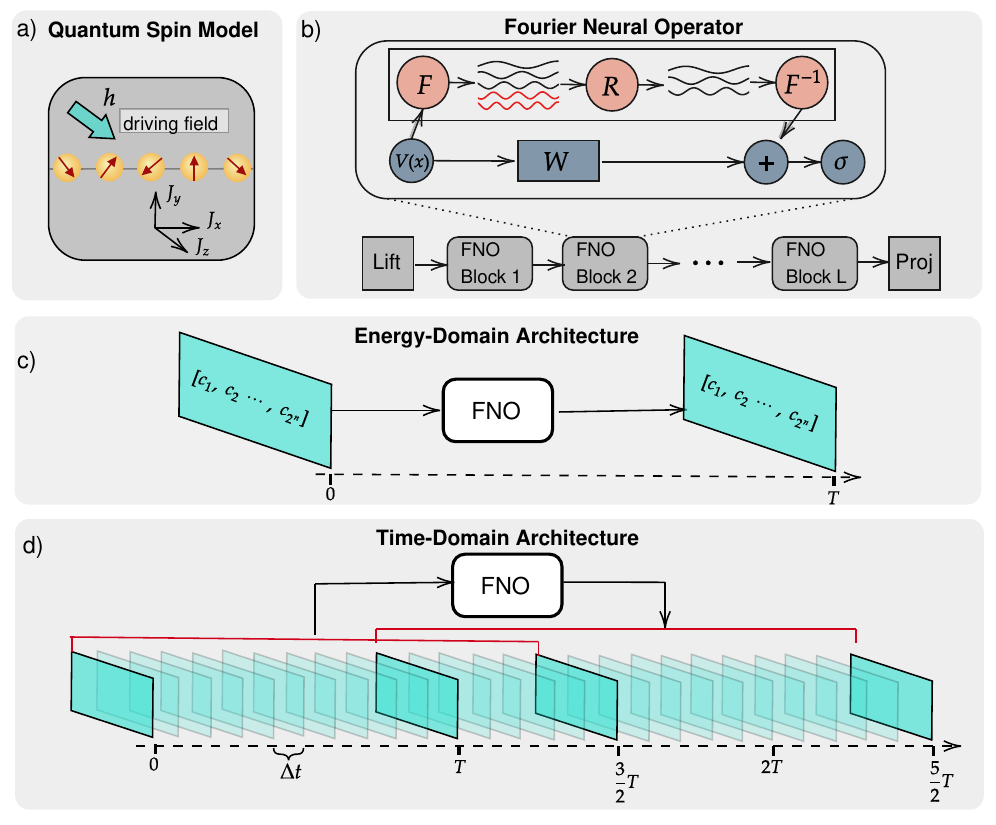}
\caption{\new{\textbf{(a)} Schematic representation of quantum spin system used to model its dynamics using FNOs, where $J_x$, $J_y$, and $J_z$ represent the coupling constants and $h$ denotes the external driving field. \textbf{(b)} Diagram illustrating the FNO framework, which consists of multiple FNO blocks performing spectral convolution over the input in the latent space. \textbf{(c)} We investigate two distinct architectures for learning dynamics in quantum spin systems. The first is the energy-domain architecture (see Section ~\ref{subsection:energy}), where the input is the wavefunction at the initial time $t=0$, and the output is the wavefunction evolved to time $t=T$.  \textbf{(d)} The second is the time-domain architecture (see Section ~\ref{subsection:time}), which takes the wavefunction evolved over the initial time interval $[0,\frac{3}{2}T]$ (discretized on a grid with width $\Delta t$) as input and produces an output wavefunction over the time interval $[T,\frac{5}{2}T]$.}}
\label{fig:model}
\end{figure*}

Moreover, traditional neural networks struggle with generalizing across different discretizations, often requiring re-training or adjustments to maintain accuracy when applied to new discretization schemes. In contrast, Fourier Neural Operators (FNOs) provide a compelling alternative by learning operators between infinite-dimensional function spaces, leveraging their resolution-invariance~\cite{li2020fourier,kossaifi2023multi,li2023fourier,lingsch2023beyond,li2024geometry,lanthaler2024discretization}. This allows FNOs to be trained at lower resolutions and seamlessly perform evaluations at higher resolutions, a phenomenon known as zero-shot super-resolution. Moreover, FNOs maintain consistent error rates across varying resolutions and offer exceptional computational efficiency, performing orders of magnitude faster than traditional solvers for partial differential equations (PDEs)~\cite{li2020fourier,maust2022fourier,bonev2023spherical,helwig2023group,zhijie2022fourier}. Recent studies demonstrate the effectiveness of FNOs in representing the S-matrix and solving fundamental quantum problems, such as the double-slit experiment and wave packet scattering \cite{mizera2023scattering}. However, the study primarily focuses on relatively simple quantum problems, which, while illustrative, have limited practical application.

\textbf{Our approach:} We explore whether FNOs can effectively learn the dynamics of quantum wavefunctions in quantum systems, such as quantum spin systems \cite{parkinson2010introduction}. Given the limitations of contemporary quantum simulators, FNOs present a potential tool for studying quantum systems, learning their dynamics and, most interestingly, extrapolating those dynamics to timescales that are not possible either experimentally (due to coherence times) or computationally (due to, e.g., the growing rank of tensor networks with increased circuit-depth or time-evolution). As quantum dynamics are dictated by the Schrödinger equation, which is a PDE, the use of FNOs for learning the time-evolution operator is well motivated. \new{We employ temporal data generated from high-fidelity classical simulations of quantum dynamics to train FNOs, enabling them to capture system evolution, extrapolate to future time intervals, and generalize for analogous inputs. This classical-data-driven approach ensures that the model learns accurate dynamics in a noise-free setting, while providing a foundation for future extensions that could use data directly from actual quantum computations to extrapolate dynamics beyond the coherence limit.} This methodology has the potential to address the limitations imposed by quantum decoherence, which impedes large-scale and prolonged simulations.

The time-evolution of spin systems is particularly advantageous for exploring the application of FNOs, as they are fundamental models that encapsulate a wide array of complex phenomena inherent in quantum mechanics \cite{sachdev1999quantum}. These simplified systems are instrumental in elucidating various quantum many-body effects, such as quantum phase transitions, gauge symmetries, and spin liquids, while also facilitating the discovery of novel, uncharacterized phenomena \cite{sandvik2010computational}. Notable examples of these models include the Ising model, the Heisenberg model, and the XY model. The large-scale simulation of these spin systems is of paramount importance due to their extensive applications in fields such as condensed matter physics, high-energy particle physics, and quantum gravity \cite{sandvik2010computational}.

In this manuscript, we numerically demonstrate the effectiveness of FNOs in learning the time evolution operator for the wavefunction of an 8-qubit Heisenberg 1D chain with random single-qubit driving. Two distinct FNO architectures are designed: the energy-domain architecture and the time-domain architecture. Moreover, we illustrate the model's ability to extrapolate dynamics to future time intervals with minimal error rates.

In addition to using these two architectures on the full quantum wavefunction, we also deploy the time-domain architecture using a mere polynomial number of Hamiltonian observables as inputs and outputs. We are motivated to study this compact architecture because the full wavefunction analyses have utility only at scales for which classical simulations can provide a ground-truth calculation. Conversely, the more compact Hamiltonian observable training procedure can be applied to data from quantum devices that exceed the capabilities of classical simulations, representing a powerful future goal. This could enable the study of quantum systems that are too large to simulate classically on timescales that are too long to carry out experimentally. Another advantage of using Hamiltonian observables as inputs is that this can push the boundaries of tensor network simulation beyond what computationally-tractable bond dimensions, which increase linearly with the number of qubits but grow exponentially with circuit or unitary depth, can allow. This is because an FNO can be trained on data from shallower tensor networks and then be used to extrapolate to longer timescales, whereas simulating the same system directly with a deeper tensor network would result in computationally prohibitive bond dimensions. To demonstrate the potential of this goal, we extrapolate the dynamics of these Hamiltonian observables into timescales longer than those provided by the training data. \new{This includes extending the Hamiltonian observable time horizon for a $20$-qubit system up to twice the duration of the training period, with a relative error of just $5.8\%$.}  

In the time-domain architecture with wavefunction inputs, we achieve a substantial 6.71x speedup with FNOs compared to exact unitary evolution for inferring dynamics at later times for 8 qubit systems with only a minimal fidelity reduction of $0.04\%$. \new{For Hamiltonian-observable inputs, the FNO demonstrates an even more remarkable inference speedup on the order of $10^4$x for 20-qubit systems, highlighting its efficiency for larger-scale quantum simulations.} We note that this speedup is likely to become more substantial at larger system sizes, as both exact unitary integration and approximate integration techniques become more computationally intensive. Additionally, we demonstrate FNO's capability for zero-shot super-resolution on 4 qubits by making predictions on a grid 10 times finer than that of the training interval. \new{Due to its discretization-invariance, the FNO maintains exceptional accuracy, achieving error rates as low as 0.04\% on the finer grid. For comparison, we include a U-Net baseline, a widely used convolutional encoder–decoder architecture with skip connections \cite{williams2023unified}. Due to its fixed-size kernels, U-Net learns patterns at a fixed resolution and does not learn the underlying operator \cite{liuschiaffini2024}. When evaluated on the same finer discretization, the U-Net exhibits a substantially higher error rate of 51.70\%. This contrast highlights that while U-Net can reproduce patterns at fixed resolution, it does not learn the underlying operator, emphasizing the advantage of FNOs for capturing system dynamics and enabling accurate extrapolation. As shown in Figure ~\ref{fig:time}, the FNO markedly outperforms the U-Net in the extrapolation regime. The U-Net performs poorly, particularly on low-energy wavefunction inputs, with a mean fidelity of $0.1501$, whereas the FNO captures the underlying structure and extrapolates with high accuracy, achieving a mean fidelity of $0.8893$.} 
\new{Similarly, a direct comparison with classical trajectory-based methods such as Dynamical Mode Decomposition (DMD) is not feasible. DMD assumes knowledge of the system’s initial state and operates on snapshot pairs from a single trajectory, learning a linear operator that describes the evolution of that specific state \cite{Zhang2016Evaluating}. In contrast, in our setting, the initial state is unknown, so DMD cannot be directly applied. The FNO, on the other hand, is trained on a diverse set of randomly generated quantum wavefunctions and learns a functional mapping of the time evolution across an entire class of states. This allows FNOs not only to reproduce local features but also to capture the underlying operator governing quantum dynamics, a capability that neither purely pattern-based ML models nor single-trajectory linear methods can provide.} These results highlight FNO’s superior performance in both computational efficiency and accuracy, emphasizing its potential as a powerful tool for predicting quantum dynamics.


\section{Results}

\subsection{Preliminaries}
In this section, we provide a brief overview of the quantum spin systems studied and outline the functionality of FNOs.
\subsubsection{Spin System Model}
\label{subsection:model}
Quantum spin systems are characterized by two-level particles organized in a specific geometry \cite{sachdev1999quantum, sandvik2010computational}. The Hamiltonian involves the interaction between neighboring quantum spins and external fields. We consider a spin $1/2$ 1D Heisenberg chain. The corresponding Hamiltonian is given by
\begin{align}
\label{equation:heisenberg}
    H &= \sum_{i=1}^{n} (J_{z}\sigma_i^{z}\sigma_{i+1}^{z}+
    J_{x}\sigma_i^{x}\sigma_{i+1}^{x}+
    J_{y}\sigma_i^{y}\sigma_{i+1}^{y}) + h\sigma_i^{z},
\end{align}
where $n$ represents the total number of qubits or atoms in the system. The Pauli matrix $\sigma_i^a$, where $a \in {x,y,z}$, is defined as $\sigma_i^a = I^{\otimes i-1}\otimes \sigma^a \otimes I^{\otimes n-i}$. Here, $I$ is the $2\times 2$ identity matrix and $\sigma^a$ denotes the corresponding $2 \times 2$ Pauli matrix. The parameters $J_{x}$, $J_{y}$, and $J_{z}$ are the coupling constants for two-qubit spin interactions, while $h$ denotes the single-qubit driving field acting along the $z$-direction. We restrict interactions to nearest neighbors and apply periodic boundary conditions, such that $\sigma_{n+1}^{a} \equiv \sigma_{1}^{a}$. For our analysis, we use a Hamiltonian with randomly assigned coupling constants and a single-qubit driving field, where the values are uniformly distributed in the range from $-2$ to $2$. Additionally, we consider the quantum Ising Hamiltonian defined as
\begin{align}
\label{equation:ising}
    H &= \sum_{i=1}^{n} J_{z}(\sigma_i^{z}\sigma_{i+1}^{z}) + h\sigma_i^{x}.
\end{align}
\\
This Hamiltonian also involves nearest-neighbor interactions and periodic boundary conditions, along with randomly assigned $J_z$ and $h$.
\subsubsection{Fourier Neural Operators}
\label{subsection:FNO}
Neural Operators (NOs) are a class of machine learning models designed to learn mappings between infinite-dimensional function spaces, making them well-suited for a variety of applications, including ordinary differential equations (ODEs) and  PDEs \cite{li2020neural,raonic2024convolutional,kovachki2023neural,azizzadenesheli2024neural}. A key advantage of NOs is their resolution-agnostic nature; they can be trained on data at one resolution and generalize to different resolutions without requiring retraining. Among Neural Operators, FNOs represent a specific implementation where spectral convolutions are utilized to capture the underlying patterns in the data~\cite{li2020fourier, FNOPracticalReview}. \new{Although the input and output are represented on discrete grids during training, the Fourier Neural Operator (FNO) learns a mapping between functions rather than between fixed-length vectors. The lifting and projection layers act pointwise with shared parameters, while the Fourier layers capture global structure in the Fourier space. Together, these components define a continuous, grid-independent operator through its Fourier representation. As a result, the trained operator can be evaluated on discretizations finer than those used in training, enabling zero-shot super-resolution without additional training. This mechanism is closely analogous to spectral interpolation on regular grids, where samples of a periodic, band-limited function defined on an $N$-point uniform grid are transformed to Fourier space, symmetrically zero-padded to a larger spectral resolution $M>N$, and then transformed back via an inverse Fourier transform to obtain the function evaluated on a $M$-point finer grid. This procedure preserves the physically meaningful low-frequency modes without introducing any artificial high-frequency modes, resulting in the evaluation of the same underlying function on a finer grid \cite{FNOPracticalReview}. This mechanism underlies the FNO’s discretization invariance, with the operator being grid-independent and the number of training points affecting only the numerical representation, not the learned mapping.}


As shown in Figure \ref{fig:model} b), an FNO consists of $L$ FNO blocks $F_{\ell}$, a lifting layer $\text{Lift}$, and a projection layer $\text{Proj}$ of the form
\begin{equation*}
    \textbf{I} \overset{\text{Lift}}{\longrightarrow} \textbf{V}_0 \overset{F_1}{\longrightarrow} \textbf{V}_1 \overset{F_2}{\longrightarrow} \ldots \overset{F_{L-1}}{\longrightarrow} \textbf{V}_{L-1} \overset{F_L}{\longrightarrow} \textbf{V}_L \overset{\text{Proj}}{\longrightarrow}\textbf{O}.
\end{equation*}
\new{The lifting layer $\text{Lift}$ is a local transformation parameterized by a shallow fully connected neural network,
\[
\mathrm{Lift} : \mathbb{R}^{d_i} \rightarrow \mathbb{R}^{d_v},
\]
which embeds the input data $\mathbf{I} \in \mathbb{R}^{d_i}$ into a higher-dimensional latent channel space. This produces the initial latent representation
\[
\mathbf{V}_0 = \mathrm{Lift}(\mathbf{I}), \qquad \mathbf{V}_0 \in \mathbb{R}^{d_v},
\]
where typically $d_v > d_i$. Here, channels denote the dimensions of the co-domain of the functions $\textbf{V}_{\ell}$, representing distinct components or features within the latent space. The latent space is an intermediate representation where the data is abstracted into a higher-level form, capturing essential patterns and relationships. The FNO operates within this latent space by applying spectral convolutions \cite{li2020fourier,FNOPracticalReview}.}


Specifically, the latent dimension representation $\textbf{V}_{\ell+1}$ is defined as
\begin{equation}
     \textbf{V}_{\ell+1}(\vec{x}) =F_{\ell}(\textbf{V}_{\ell})(\vec{x}) =  \sigma (W_t\textbf{V}_{\ell}(\vec{x})+(K_{\ell}\textbf{V}_{\ell})(\vec{x})),
\end{equation}
where $W_{\ell}$ is a learnable affine-linear map applied across the channels of $\textbf{V}_{\ell}$, 
and $\sigma$ is a non-linear activation function. \new{For all FNO experiments in this work, we employ the Gaussian Error Linear Unit (GELU) as the activation function. We choose GELU because it has been shown to perform well in smooth operator learning tasks and is more flexible than ReLU, allowing small negative outputs and being conducive to more stable training \cite{FNOPracticalReview}.}
The spectral convolution $K_{\ell}$ can be defined as follows,
\begin{equation}
K_t\textbf{V}_{\ell} = \mathcal{F}^{-1}\left(R_{\ell} \cdot \mathcal{F}(\textbf{V}_{\ell})\right), 
\end{equation}
where $\mathcal{F}$ and $\mathcal{F}^{-1}$ denote the Fourier and inverse Fourier transforms, respectively. In the Fourier domain, higher frequency modes are truncated, leaving a fixed number of lower modes that are multiplied with learnable parameters $R_{\ell}$. By combining the power of linear transformations, spectral convolutions, and nonlinear activation functions, the FNO can approximate highly non-linear operators~\cite{lanthaler2023nonlocal}. 

\new{After processing through multiple FNO blocks $F_{\ell}$, the projection layer $\text{Proj}$ maps the latent representation back to the output function space. 
\[
\mathrm{Proj} : \mathbb{R}^{d_v} \rightarrow \mathbb{R}^{d_o},
\]
which yields the predicted output
\[
\mathbf{O} = \text{Proj}(\mathbf{V}_L), \qquad \mathbf{O} \in \mathbb{R}^{d_o}.
\]
 The output data $\textbf{O}$ may have one or more channels, depending on the specific architecture used as described in Sections \ref{subsection:energy} and \ref{subsection:time}.}

\subsection{Learning Dynamics using Complete $2^n$ Wavefunction}
\label{Section:1B}
\new{In this approach, the full state space of a quantum wavefunction is used as input, with two distinct types of wavefunctions considered. The first type consists of complex-valued normalized wavefunctions for \(n\) particles. Each amplitude is generated by independently sampling its real and imaginary parts from a uniform distribution on the interval \([0,1)\) using \texttt{torch.rand}. The resulting wavefunction is then normalized to have unit \(L^2\) norm.} For the second type, we uniformly distribute the wavefunction over low-energy states while setting the high-energy states to zero. \new{The exact method is described in detail in Section \ref{section:methods}.} This distinction is insightful, as in many physical applications, e.g., quantum chemistry \cite{levine2009quantum}, often only low-energy components of wavefunctions are occupied. In the remainder of this manuscript, we refer to the first input type as ``random input'' and the latter as ``low-energy input''. We develop two distinct FNO architectures for processing wavefunction inputs: the energy-domain and time-domain architectures, which are described in detail below. Notably, because quantum wavefunctions are complex quantities, we use a complex version of the FNO \cite{trabelsi2018deep}.
\subsubsection{Energy-domain Architecture}
\label{subsection:energy}

We first consider the energy-domain architecture, where the Fourier transform is applied to the basis states of the wavefunction. This architecture requires a careful ordering of the wavefunction, specifically by arranging the basis states in order of increasing energy levels, so that states with lower energies precede those with higher energies. The method is described in detail in Section \ref{section:methods}. In the Fourier domain, the FNO truncates fast (high-frequency) energy transitions. In quantum physics, energy $E$ and frequency $\nu$ are directly related by the Planck-Einstein relation $E=h\nu$, where $h$ is the Plank's constant. Therefore, performing a Fourier transform over energy is effectively the same as performing it over frequency.

In this architecture, the input comprises various training data of the quantum wavefunction at an initial time, as shown in Figure \ref{fig:model} c). The input to the FNO is structured as follows
\begin{equation}
    \textbf{I}=\big[\operatorname{Embed_{\mathrm{state}}},S_{0}\big] \in \mathbb{C}^{2 \times 2^{n}}.
\end{equation}
\new{Here, $S_0$ is the vector of complex amplitudes of the $2^n$ wavefunction in the computational basis at initial time $t=0$, $S_0 = \left[ c_1, c_2, \dots, c_{2^n} \right]$, with $\sum_{i=1}^{2^n} |c_i|^2 = 1.$} To enhance the learning process, we incorporate a position embedding 
$\operatorname{Embed_{\mathrm{state}}}(k)=k/2^n$ into the input channel dimension, where $k$ is the index associated with each basis state. \new{In this setting, the ordering of the basis states corresponds to increasing energy levels. By providing this positional embedding, the FNO can more effectively capture the hierarchical structure of the wavefunction, distinguishing between low- and high-energy components. This is particularly relevant because the FNO operates in Fourier space, where high-frequency modes are truncated. The position embedding ensures that these truncated modes predominantly correspond to high-energy components that contribute minimally to the system’s dynamics, thereby enhancing both the learning efficiency and extrapolation accuracy.} 
The output tensor is given by
\begin{equation}
    \textbf{O} = S_{T}  \in \mathbb{C}^{2^n}.
\end{equation}
Thus the output, 
consists of wavefunction output evolved at $t=T$. In our experiments, we choose $T=\pi$ as it represents a significant portion of the quasi-periodicity of the unitary evolution generated by the Hamiltonian $H$. \new{The Fourier transform on the amplitudes $c_k$ of the wavefunction in the computational basis is given by
\begin{equation} 
    \frac{1}{2^n}\sum_{k=0}^{2^n-1} c_{k} e^{\frac{-i k \tau}{\hbar}},
\end{equation}
where $c_k$ is the complex amplitude of the $k$-th computational basis state, $\tau$ can be understood as a time-like variable, and $\hbar$ is the reduced Planck constant.
} By truncating the transformed modes, we effectively filter out the components where energy values change rapidly, thus eliminating fast transitions. By concentrating on the slow-transition dynamics, the FNO highlights the most relevant dynamics of the quantum system. While this architecture provides a physical interpretation as a Fourier transform along the wavefunction's energy space, it does not support further discretization of the time interval $[0,T]$, thereby limiting its utility in scenarios requiring more fine-grained prediction of the time-evolution. We will address this issue with our time-domain architecture in Section~\ref{subsection:time}.

            
\begin{figure*}[t]
    \begin{subfigure}[t]{0.55\textwidth}
        \raggedright
        \textbf{(a)}\par
        \includegraphics[width=\textwidth]{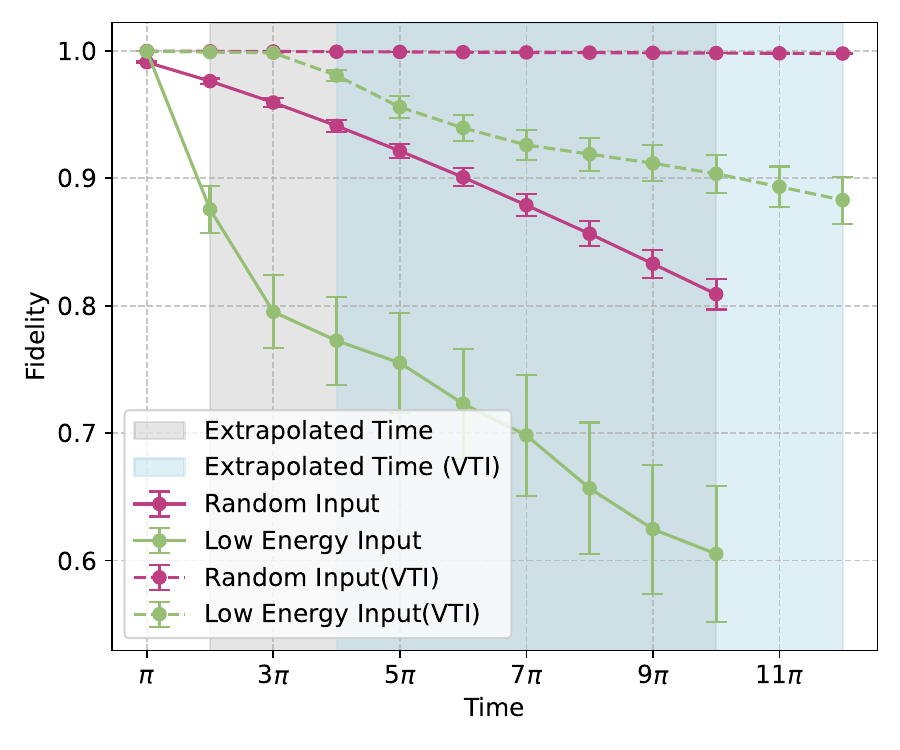} 
    \end{subfigure} 
    \begin{subfigure}[t]{0.5\textwidth}
        \raggedright
        \textbf{(b)}\par
        \centering
        \textbf{Random initial states}
        \vspace{0.4em}
    
        \setlength{\tabcolsep}{8pt}
        \makebox[\linewidth][c]{
        \begin{tabular}{c c c c c}
            \hline
            \textbf{n} & $F(\pi)^{\mathrm{train}}$ & $F(2\pi)^{\mathrm{ext}}$ & $F(6\pi)^{\mathrm{ext}}$ & $F(10\pi)^{\mathrm{ext}}$ \\
            \hline
            4 &
            $0.9997 \pm 0.0001$ &
            $0.9991 \pm 0.0003$ &
            $0.9949 \pm 0.0026$ &
            $0.9894 \pm 0.0060$ \\
            8 & 
            $0.9911 \pm 0.0009$ & 
            $0.9762 \pm 0.0023$ & 
            $0.9010 \pm 0.0074$ & 
            $0.8101 \pm 0.0129$ \\
            \hline
        \end{tabular}
        }
    
        \vspace{1em}
    
        \textbf{Low-energy states (VTI)}
        \vspace{0.5em}
    
        \setlength{\tabcolsep}{8pt}
        \makebox[\linewidth][c]{
        \begin{tabular}{c c c c c}
            \hline
            \textbf{n} & $F(3\pi)^{\mathrm{train}}$ & $F(4\pi)^{\mathrm{ext}}$ & $F(8\pi)^{\mathrm{ext}}$ & $F(12\pi)^{\mathrm{ext}}$ \\
            \hline
            4 &
            $0.9999 \pm 0.0001$ &
            $0.9999 \pm 0.0001$ &
            $0.9994 \pm 0.0005$ &
            $0.9984 \pm 0.0014$ \\
            8 & 
            $0.9984 \pm 0.0002$&
            $0.9806 \pm 0.0047$& 
            $0.9202 \pm 0.0125$& 
            $0.8836 \pm 0.0184$\\
            \hline
        \end{tabular}
        }
    \end{subfigure}

    \caption{\textbf{(a)} Prediction of temporal dynamics in an 8-qubit Heisenberg spin system using the energy-domain architecture in Section~\ref{subsection:energy}. At time $T=\pi$, the predictions made by the FNO are on unseen data but still within the time range it was trained on. The extrapolated time refers to future time predictions beyond the range on which the FNO was trained. VTI (Various Time Interval) indicates training on multiple time intervals, i.e., $[0, T]$, $[T, 2T]$, and $[2T, 3T]$, instead of training only on the first interval. \textbf{(b)} The results show the fidelity at specific time steps for 4 and 8 qubits, including times within the training range as well as extrapolated future times beyond the training interval. Superscripts indicate whether the particular time step is within the training range ($\mathrm{train}$) or an extrapolated time ($\mathrm{ext}$).}
    \label{fig:energy_transformed}
\end{figure*}

We use fidelity as the primary metric to evaluate the performance of the FNO in predicting quantum wavefunction dynamics. Fidelity measures the similarity between two quantum states and is defined as,
\begin{equation}
    F(\psi_{\text{true}}, \psi_{\text{pred}}) = \left| \langle \psi_{\text{true}} | \psi_{\text{pred}} \rangle \right|^2,
\end{equation}
where \(\langle \psi_{\text{true}} | \psi_{\text{pred}} \rangle\) is the inner product between the true and predicted wavefunctions. Additionally, we iteratively apply the model to its own predictions to forecast wavefunction dynamics over a period of up to $t=10T$. Although the model is trained on the time interval $[0,T]$, this iterative approach allows it to predict wavefunction evolution well beyond the training data time-horizon.

Experiments are conducted using $4$ and $8$ qubits with both random and low-energy input states. For low-energy states, training is conducted over various time intervals (VTI), such as $[0,T]$, $[T,2T]$, and $[2T,3T]$, rather than relying on a single time interval $[0,T]$. Utilizing multiple time intervals is crucial to generalize to future time intervals for low-energy states. Specifically, when training on only a single time step, the model fails to adequately capture the dynamics necessary for accurate future time predictions, despite showing strong performance during training. This is seen in Figure \ref{fig:energy_transformed}, where 8-qubit with low energy states has a mean fidelity of $0.7538$ for prediction of future time until $t=10T$  despite achieving a fidelity of $0.9998$ during training. This discrepancy underscores the importance of multi-step training for low-energy states, as these states initially occupy only low-energy configurations and gradually transition to higher-energy states over time. \new{When VTI is applied, low-energy inputs achieve substantially higher mean fidelities at extrapolated times, reaching $0.9992$ for 4-qubit and $0.9430$ for 8-qubit states. We also test VTI when training models on random inputs. Random inputs do not strictly require multiple intervals, as they already occupy a wide array of energy regions and can successfully extrapolate to later times. For example, 4-qubit random inputs without VTI reach a mean fidelity of about $97\%$, while 8-qubit random inputs achieve around $91\%$. Nonetheless, applying VTI even for random inputs further improves performance, achieving mean fidelities as high as $99.99\%$ for 8 qubits, highlighting that multi-interval training can still be beneficial.} \\ \new{FNO performs better on random wavefunctions as these states already span a wide range of energy modes and configurations. Each random state effectively provides the model with information across the entire Hilbert space, allowing it to learn the dynamics for all relevant energy regions in a single step. In contrast, low-energy states are initially localized to only a small subset of modes, and accurately predicting their future evolution requires the model to capture how population spreads to higher-energy modes over time. This highlights why VTI is necessary for accurately extrapolating low-energy states, whereas random states achieve satisfactory results even when the model is trained on a single time step.} Figure \ref{fig:energy_transformed} demonstrates that the model achieves high fidelity for both input types and can accurately predict future states while maintaining consistent fidelity across these predictions.

\subsubsection{Time-domain Architecture}
\label{subsection:time}

\begin{figure*}[htb]
\centering
\includegraphics[width=\textwidth]{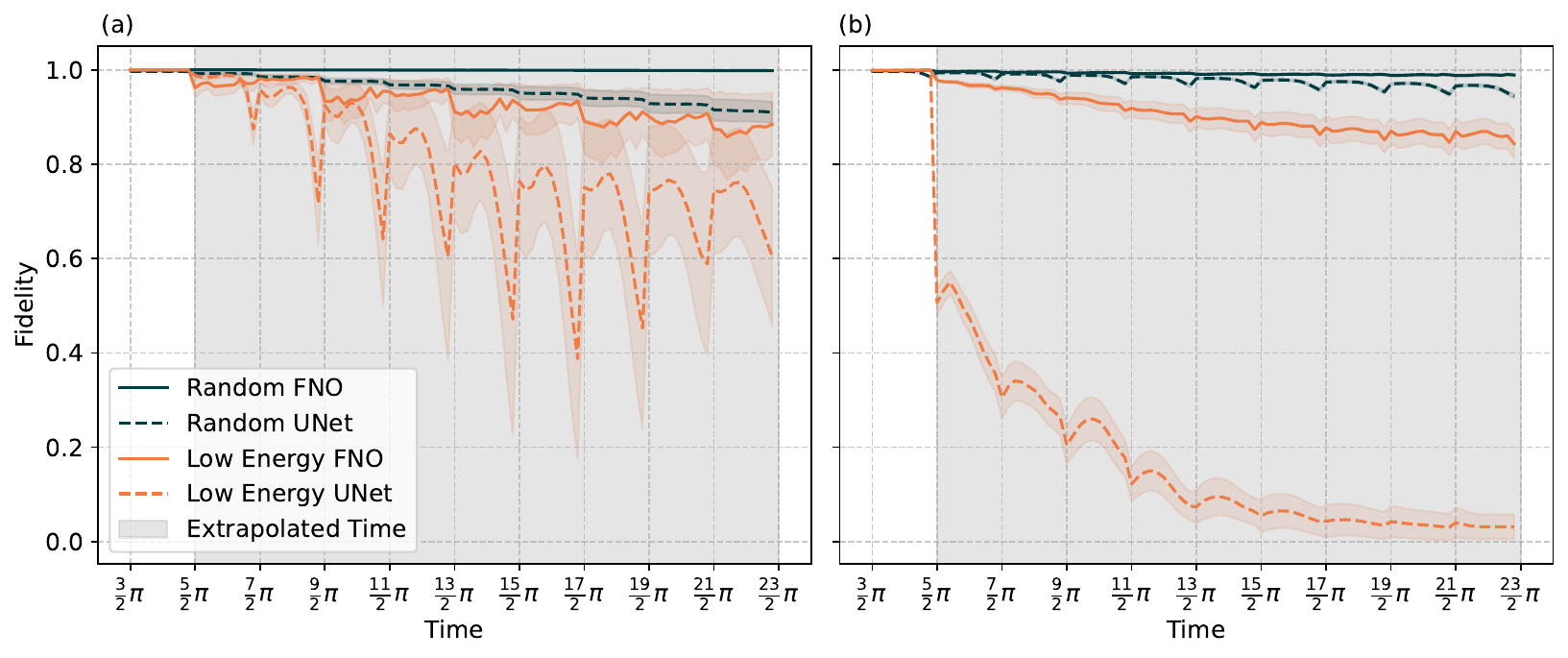}
    \caption{\new{\textbf{(a)} Prediction of temporal dynamics in a four-qubit Heisenberg spin system using the time-domain architecture in Section \ref{subsection:time}. The model is trained on the time interval $[0, 3\pi/2]$ as illustrated in Figure \ref{fig:model}d, while the extrapolated time refers to future predictions beyond the trained range. Performance for both the random and low-energy wavefunction initial states using the FNO is benchmarked against a U-Net baseline. The shaded regions around each curve indicate one standard deviation. The FNO accurately captures the underlying operator and extrapolates reliably to later times, while the U-Net exhibits poor extrapolation, especially for low-energy input states. \textbf{(b)} Corresponding results for an eight-qubit Heisenberg spin system.}}
\label{fig:time}
\end{figure*}

An alternative architecture involves predicting the evolution of the wavefunction on a whole time interval instead of a single time $T$. In practice, we need to discretize the time interval. However, using a discretization-agnostic model, such as an FNO, we train and predict using grids with arbitrary width $\Delta t$. Instead of a single time $t=0$, we then use a time interval as input, i.e.,
\begin{equation}
\label{equation:time input}
    \textbf{I}=\big[\operatorname{Embed_{\mathrm{time}}},S_{[0,\frac{3}{2}T]} \big] \in \mathbb{C}^{ (2^{n}+2) \times m },
\end{equation}
where $m=\frac{3}{2}T / \Delta t$ defines the number of of equidistant time-steps in the interval $[0,\frac{3}{2}T]$. For training, we choose $T=\pi$ and $\Delta t=\pi/10$.
\new{For the positional embedding $\operatorname{Embed}_{\mathrm{time}}$, we use a sinusoidal encoding to facilitate the FNO in learning temporal patterns. Given a discretized time grid with $m$ equidistant points, indexed by $l \in \{0,\dots,m-1\}$, the positional embedding is constructed using a single cosine–sine pair. With two embedding channels and a fixed frequency $\omega = 1$, each time index $l$ is mapped as
\begin{equation}
\operatorname{Embed}_{\mathrm{time}}(l)
=
\big[ \cos(\omega l), \; \sin(\omega l) \big].
\end{equation}
This results in a positional embedding tensor of shape $(1,2,m)$, where the two channels correspond to the cosine and sine components evaluated at each time step. By mapping time onto a smooth periodic representation, the embedding allows the model to capture the relative ordering of temporal points while preserving continuity across the discretized interval \cite{li2020fourier,FNOPracticalReview}. Although the approach can be generalized to multiple frequencies by increasing the number of channels, in this work we use the minimal two-channel formulation.} The output of the FNO is given as
\begin{equation}
    \textbf{O} = S_{[T, \frac{5}{2}T]}\in \mathbb{C}^{2^n \times m},
\end{equation}
\noindent 
where $S_{[T, \frac{5}{2}T]}$ denotes the wavefunction evolved over the (discretized) time interval $[T, \frac{5}{2}T]$, where we overlap the time interval $[T, \frac{3}{2}T]$ with the one of the input $\textbf{I}$ in Eq.~\eqref{equation:time input}.
This setup facilitates smoother and more accurate learning, as the FNO can leverage the temporal continuity and patterns in the evolved wavefunction data. Consequently, the output includes the future time interval $[\frac{3}{2}T,\frac{5}{2}T]$, unseen during training. We use two different sets of previously described inputs, i.e., random wavefunctions and low-energy wavefunctions. While the basis states does not necessarily need to be ordered as in the previous architecture, we maintain the order for consistency.

We also perform extrapolated time prediction by applying the model on its predicted time intervals to predict unseen future time intervals, up to $t=\frac{23}{2}T$, as seen in Figure \ref{fig:time}. The fidelity metrics for both the training intervals and the extrapolated times are also reported in Figure~\ref{fig:time}, demonstrating that even with substantial predictions into the future, the FNO achieves exceptionally high fidelities for systems of both 4 and 8 qubits. Additionally, we compare the performance of the FNO with that of a deep neural network, specifically a U-Net \cite{williams2023unified}. Our evaluation demonstrates that the FNO outperforms the U-Net, especially in the extrapolated time regime, as seen in Figure \ref{fig:time}. Moreover, even at moderate system sizes (8 qubits), the FNO achieves an 6.71x speedup compared to the exact unitary-based method in predicting the dynamics of random inputs up to $t=\frac{23}{2}T$, with only a negligible fidelity reduction of $0.04\%$. \new{This speedup does not take into account the training cost, which for a moderate size of 8 qubits is approximately $2$s per epoch over 200 training epochs.}

Additionally, we performed evaluations on the output time interval $[T, \frac{5}{2}T]$ using a finer grid with width $\Delta t = \pi/100$ for a 4-qubit system with random inputs. 
For the FNO, the resulting fidelity of $0.9999$ was identical to that obtained on the coarser training discretization. While we can also apply the U-Net on the finer grid, it is not discretization-agnostic, leading to an error rate of $ 4.78\%$ in the expected fidelity. To further demonstrate the FNO's capability to achieve zero-shot super-resolution, we used an input interval of $[0, \frac{3}{2}T]$ with $T=5\pi$ and $\Delta t=\pi/2$ for 4 qubits and evaluated the corresponding output interval on a finer grid with $\Delta t=\pi/20$. The FNO successfully predicted the finely discretized output with an error rate of just $0.04\%$, whereas the U-Net demonstrated a significantly higher error rate of $51.70\%$ compared to the coarse grid fidelity.

\subsection{Learning Dynamics using Hamiltonian Observables}
\label{Subsection:HO}
%

\begin{figure*}[htb]
\centering

\begin{minipage}[t]{0.76\textwidth}  
    \raggedright
    \textbf{(a)}
    \includegraphics[width=\textwidth]{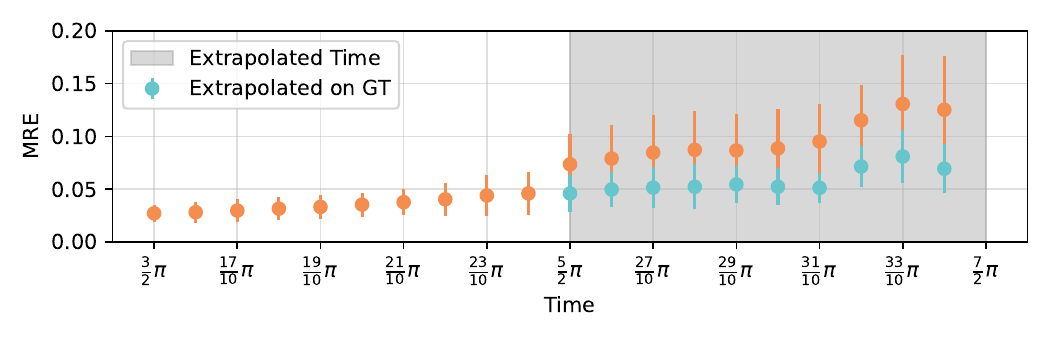}
\end{minipage}%
\hfill
\begin{minipage}[t]{0.76\textwidth}  
    \raggedright
    \textbf{(b)}\par\vspace{0.3em}
    {
    \setlength{\tabcolsep}{8pt}
    \makebox[\linewidth][c]{
    \begin{tabular}{c c c c c}
        \hline
        \textbf{n} &
        $\mathrm{MRE}(24\pi/10)$ &
        $\mathrm{MRE-GT}(5\pi/2)$ &
        $\mathrm{MRE-GT}(31\pi/10)$ &
        $\mathrm{MRE-GT}(34\pi/10)$ \\
        \hline
        8  & 0.0675 $\pm$ 0.0305 &  0.0242 $\pm$ 0.0114 & 0.0815 $\pm$ 0.0484 & 0.1057 $\pm$ 0.0573 \\
        20 & 0.0460 $\pm$ 0.0208 & 0.0459 $\pm$ 0.0176 & 0.0515 $\pm$ 0.0150 & 0.0704 $\pm$ 0.0247 \\
        \hline
    \end{tabular}
    }}
\end{minipage}

\captionsetup{justification=raggedright,singlelinecheck=false}
\caption{
\new{\textbf{(a)} Mean Relative Error (MRE) for predicted Hamiltonian observables in a 20-qubit system using FNO, utilizing the time-domain architecture. The model is trained on the input time interval $[0,\frac{3}{2}\pi]$ as shown in Figure \ref{fig:model} d) and predicts the output interval $[\frac{3}{2}\pi,\frac{5}{2}\pi]$. Additionally, it extrapolates future unseen dynamics $[\frac{5}{2}\pi,\frac{7}{2}\pi]$, which is twice the length of the output interval. The MRE is $5.8\%$ over the extrapolated time predictions on ground truth (GT). This is particularly significant in quantum simulations, where the primary objective is to extend these simulations. Additionally, based on the time interval the model was trained on, the MRE for extrapolated time predictions is $9.6\%$ denoted by orange dashes. We additionally calculate the mean MSE loss for the training time interval as \(2.16 \times 10^{-9} \pm 2.15 \times 10^{-9}\) and for future time predictions over the same interval on ground truth as \(5.34 \times 10^{-9} \pm 4.13 \times 10^{-9}\). \textbf{(b)} The results present the MRE at specific time steps for both the train time (i.e., the time intervals that FNO encountered during training) and the extrapolated time [$\frac{5}{2}\pi, \frac{7}{2}\pi$] for systems with 8 and 20 qubits for hamiltonian observables. }
}
\label{fig:observables}
\end{figure*}

\new{Hamiltonian observables refer to the individual operator terms that appear in the
time-evolution generator (the Hamiltonian) of the system. In practice, we do not use these operators themselves as inputs to the model. Rather, we compute their expectation values with respect to the evolving quantum state and use those quantities as the actual inputs. For example, in the Ising model,
the relevant operators include terms such as $\sigma_z \sigma_z$ and $\sigma_z$,
while in the Heisenberg model they include $\sigma_a \sigma_a$ for
$a \in \{x,y,z\}$ and $\sigma_z$. We evaluate the corresponding expectation values $\langle O_j \rangle_t = \langle \psi(t)|O_j|\psi(t)\rangle$ at each time step, which then form the real-valued input and output channels for the FNO. Using expectation values of Hamiltonian, instead of the full wavefunction, as inputs and outputs for the FNO greatly enhances its scalability, as there are approximately $\text{poly}(n)$ terms in the former and $2^n$ terms in the latter.} This compression would allow us to train and make predictions with quantum systems that are too large or long to simulate directly, e.g., by using data from large quantum devices or wide tensor networks, from which we could then infer times longer than the devices' coherence time or past the tensor network's tractable depth, respectively. However, predicting future dynamics based on these partial observables presents a significant challenge, particularly without explicit knowledge of the underlying Hamiltonian.

To evaluate the FNO's capabilities, we first considered the Ising Hamiltonian
as described in Eq.~\eqref{equation:ising}, where $J_z$ and $h$ are randomly generated. 
\new{This introduces variability in the Hamiltonian parameters, ensuring that the learning task is general with respect to the Hamiltonian and hence challenging. Importantly, in this study, we focus on using random wavefunction states to calculate the expectation values of these observables.}

\new{For the observable-based inputs,  we exclusively utilize only the time-domain architecture. In this configuration, the model receives a fixed set of observable expectation values over the chosen time interval. A detailed specification of the observables appears in Section \ref{section:methods}. This input structure mirrors the previously described input $\textbf{I}$ in Eq.~\eqref{equation:time input}, but instead of $2^n$ wavefunction basis states $S$, it includes a substantially smaller set of observable expectation values. We use the same input time interval, $[0, 3T/2]$, with $T = \pi$ and time step $\Delta t = \pi/10$, as described in Section~\ref{subsection:time}.} This setup presents a substantial challenge for the FNO, as it must learn to predict future dynamics based on only partial information about the quantum state. 

\new{For our experiments, we consider system sizes of 8 and 20 qubits. For the larger system (20 qubits), the time-evolved expectation-value data is generated using the \texttt{CUDA-Q} library with Trotterization, since exact simulation at these scales is computationally intractable.} We further extrapolate to the future time interval $[\frac{5}{2}T, \frac{7}{2}T]$, effectively doubling the dynamics captured within the output time interval $[\frac{3}{2}T, \frac{5}{2}T]$. This feature is particularly advantageous for quantum computing, where the limited coherence time of error-corrected qubits imposes constraints on the length of computations. Additionally, it can also be integrated to take observable data from tensor network methods and extrapolate it to timescales that would be computationally difficult for it to calculate due to high tensor train rank. Given that these observables present challenges for the FNO to learn effectively, we extrapolate future times based on the ground truth of the predicted time interval. This approach helps minimize unnecessary errors, as the ground truth is typically known. 
\new{As seen in Figure \ref{fig:observables}, the FNO maintains consistent and accurate performance even as the system size increases substantially. For 8- and 20-qubit systems, we predict future observables with relative errors of $6.46\%$ and $5.8\%$, respectively, even when extending the prediction to twice the training time horizon on ground truth data. The model exhibits no degradation in accuracy with increasing system size, and its extrapolation performance remains stable across all settings. This strongly demonstrates that FNOs can scale effectively to larger quantum many-body systems while preserving error rates in both the training regime and long-time extrapolation. It is also noteworthy that the required volume of training data does not scale in proportion to the exponential growth of the Hilbert space. For instance, we use 18,000 training samples for the 8-qubit system and only 22,500 for the 20-qubit system. This modest increase, despite the dramatic growth in system dimensionality, underscores the data efficiency of the FNO architecture and further illustrates its practical scalability for large-scale quantum dynamical prediction.} 

\new{Once trained, FNO enables extremely fast inference of quantum dynamics. For a representative example involving 1500 time steps, FNO inference requires 7.1 s in total, whereas GPU-accelerated Trotterized time evolution implemented in CUDA-Q requires approximately 85 s per time step on a Grace Hopper system and 180 s per time step on an NVIDIA A100 GPU. This corresponds to an inference-time speedup of approximately $1.8 \times 10^{4}$ (Grace Hopper) to $3.8 \times 10^{4}$ (A100) compared to numerical time-evolution methods. Although this does not include the training cost of the FNO, this cost is incurred only once. For the largest system considered (20 qubits), training requires approximately 40 s per epoch over 1700 epochs, corresponding to a total training time of about 19 hours using an A100 GPU. This cost is substantially reduced for smaller systems, such as 8 qubits, where the training time per epoch is approximately 22 s. While training the FNO entails a substantial one-time computational cost that grows with system size, this cost can be amortized across multiple subsequent inference tasks, such as computing dynamics for new initial states. In contrast, numerical time-evolution methods must be rerun independently for each such trajectory, resulting in a computational cost that grows linearly with the number of simulations.}

\section{Discussions}
\label{section:discussions}
We have presented two FNO architectures that are capable of not only learning the time-evolutions of quantum systems, but also of extrapolating these evolutions to later times. The energy-domain architecture is more compact, requiring fewer computational inputs, and hones in on key dynamics by prioritizing slow quantum transitions.  The time-domain architecture is agnostic to the discretization of the time interval, making it highly versatile for obtaining the output state of quantum evolution at arbitrary times. \new{However, this approach does not scale efficiently, since the Hilbert space dimension of the wavefunction increases exponentially with the number of qubits. Nevertheless, it remains useful for faster evaluation as, for an 8-qubit system, it yields a $6.71$x speedup in evaluation compared to exact solvers. Furthermore, because the FNO learns the underlying operator and considers the full wavefunction as input, it can reliably extrapolate to approximately $10$x the training time, a regime that is computationally demanding for exact solvers and inaccessible to tensor-network methods due due to increasing entanglement.} Both methods demonstrate the ability of FNOs to not only carry out quantum state evolution, but to learn the underlying time-evolution operator itself, which constitutes a key accomplishment in the ubiquitous and challenging task of solving the Schrödinger equation.

Of foremost interest is the application of the time-domain FNO using only Hamiltonian observables as input and output. This difficult learning task requires that the FNO learn and conduct future inference on partial information, using only $\sim\text{poly}(n)$ expectation values rather than the full $2^n$-component wavefunction. \new{Notably, our results demonstrate that the FNO maintains consistent performance as system size increases substantially. As shown in Figure \ref{fig:observables}, for 8- and 20-qubit systems, the FNO predicts future observables with relative errors of 6.46\% and 5.8\%, respectively, even when extrapolating to twice the training time horizon on ground truth data. Furthermore, it requires only a modest increase in the number of training samples to scale efficiently to larger quantum many-body systems.}

\new{These results establish FNO as a promising computational framework for exploring novel quantum phenomena and simulating complex quantum many-body systems. Its favorable scalability, capacity for long-time extrapolation, and data-efficient learning enable access to regimes that are challenging for conventional numerical approaches. In particular, the FNO could be explored as a data-driven framework to learn the dynamics of a broad class of systems, including interacting spin chains, fermionic lattice models, and disordered quantum systems that are otherwise computationally intractable. Within this architecture, the FNO may be trained on measurement data obtained from noisy intermediate-scale quantum devices or on data generated by shallow tensor-network simulations, and subsequently leveraged to predict system dynamics at later time scales. Such predictions would otherwise necessitate either highly coherent quantum hardware or tensor-network representations of prohibitive bond dimension.} 

In subsequent research, such a hybrid implementation of our work should be carried out using a large quantum device and a classical FNO to extrapolate to longer times. As open quantum systems have distinct dynamics from their pure counterparts, FNOs should also be applied to noisy quantum states. Moreover, the impact of physical noise and system symmetry on the requisite FNO dimensions and training data size should be studied. As quantum noise, sampling errors, and symmetries can reduce the learning complexity of the quantum system, it is natural that we characterize the FNO in this capacity.

\section{Methods}
\label{section:methods}
We provide details on the low-energy states generation in Section \ref{Section:1B}, wavefunction ordering protocol used in Section \ref{subsection:energy}, the types of Pauli strings utilized in Section \ref{Subsection:HO}, and the specifics of the training data, and FNO configurations for Figures \ref{fig:energy_transformed}, \ref{fig:time}, and \ref{fig:observables}.

\new{We generate the low-energy states referenced in Section \ref{Section:1B} by computing the energies of all $2^n$ computational-basis states under the system Hamiltonian and select the $M$ configurations with the lowest energies. A low-energy wavefunction is constructed by assigning complex amplitudes only to these $M$ modes, sampled
independently from a uniform distribution, while all remaining basis-state amplitudes
are set to zero. The resulting state is normalized to have unit $L^2$ norm. In our
numerical settings, we use $M = 4$ for the 4-qubit system and $M = 50$ for the
8-qubit system.}

The quantum wavefunction $\psi$ of a system with $n$ qubits is represented as a vector in a $2^n$-dimensional Hilbert space. The basis states $\phi_i$ correspond to different qubit configurations, ordered by their binary representation. To reorder the wavefunction by energy levels, we arrange the basis states such that their associated energies $E_i$ satisfy $E_i \leq E_{i+1}$. The wavefunction is then expressed as,

\begin{equation}
    \psi = \sum_{i=1}^{2^n} c_i \phi_i.
\end{equation}

Here $\phi_i$ are ordered according to increasing energy levels. The energy levels can be calculated using $E_i = \langle \phi_i | H | \phi_i \rangle$, where $H$ is the Hamiltonian of the system.

In Section \ref{Subsection:HO}, for a system of 8 qubits, we use a set of 48 observables that includes all nearest-neighbor interactions $XX$, $YY$, and $ZZ$ interactions across all qubit pairs. Additionally, the set includes single-qubit interactions $X$, $Y$ and $Z$. We focus exclusively on quantities exceeding a certain threshold (e.g., greater than $10^{-2}$) to avoid including less significant values that might inflate the relative error. Given that these observables are the expectation values of Pauli strings, we employ Mean Squared Error (MSE) and Mean Relative Error (MRE) as our loss metrics to evaluate the model’s performance.

In Figure \ref{fig:energy_transformed}, an 8-qubit system is used with 4,000 training data points, 4 FNO blocks, and 128 modes retained in the Fourier integral operator after truncation. For low-energy states with VTI, each interval contains 5,000 training data points.

In Figure \ref{fig:time}, the model is again trained with 4,000 data points, using 4 layers and retaining 7 modes in the FNO after truncation. The U-Net is also trained on the same amount of data.

In Figure \ref{fig:observables}, we use 18,000 training data points with 48 Pauli strings for the 8-qubit system, utilizing 4 layers and retaining 7 modes. \new{For 20 qubits, we use 22500 training data points with 120 Pauli strings, 4 layers, and retaining 7 modes.}

\section{acknowledgments}
F.S.\@ acknowledges support from the Caltech Summer Undergraduate Fellowship. J.B.\@ acknowledges support from the Wally Baer and Jeri Weiss Postdoctoral Fellowship. A.A.'s work is supported in part by the Bren endowed chair, the ONR (MURI grant N00014-18-12624), and the AI2050 Senior Fellow Program at Schmidt Sciences. We acknowledge the Amazon Web Services (AWS) Advanced Compute team, with particular thanks to Tyler Y. Takeshita and Sebastian Hassinger, for their compute support.

\section*{Code availability}
The code to reproduce the results reported in this study is available from the corresponding author upon request.


\bibliographystyle{apsrev4-2}
\bibliography{reference}

@incollection{panagakis2024tensor,
  title={Tensor methods in deep learning},
  author={Panagakis, Yannis and Kossaifi, Jean and Chrysos, Grigorios G and Oldfield, James and Patti, Taylor and Nicolaou, Mihalis A and Anandkumar, Anima and Zafeiriou, Stefanos},
  booktitle={Signal Processing and Machine Learning Theory},
  pages={1009--1048},
  year={2024},
  publisher={Elsevier}
}

@article{kovachki2023neural,
  title={Neural operator: Learning maps between function spaces with applications to {PDEs}},
  author={Kovachki, Nikola and Li, Zongyi and Liu, Burigede and Azizzadenesheli, Kamyar and Bhattacharya, Kaushik and Stuart, Andrew and Anandkumar, Anima},
  journal={Journal of Machine Learning Research},
  volume={24},
  number={89},
  pages={1--97},
  year={2023}
}

@article{lingsch2023beyond,
  title={Beyond Regular Grids: {Fourier}-Based Neural Operators on Arbitrary Domains},
  author={Lingsch, Levi and Michelis, Mike Y and de Bezenac, Emmanuel and Perera, Sirani M and Katzschmann, Robert K and Mishra, Siddhartha},
  journal={arXiv preprint arXiv:2305.19663},
  year={2023}
}

@article{maust2022fourier,
  title={Fourier continuation for exact derivative computation in physics-informed neural operators},
  author={Maust, Haydn and Li, Zongyi and Wang, Yixuan and Leibovici, Daniel and Bruno, Oscar and Hou, Thomas and Anandkumar, Anima},
  journal={arXiv preprint arXiv:2211.15960},
  year={2022}
}

@inproceedings{raonic2024convolutional,
  title={Convolutional neural operators for robust and accurate learning of {PDEs}},
  author={Raonic, Bogdan and Molinaro, Roberto and De Ryck, Tim and Rohner, Tobias and Bartolucci, Francesca and Alaifari, Rima and Mishra, Siddhartha and de B{\'e}zenac, Emmanuel},
  booktitle={Advances in Neural Information Processing Systems},
  volume={36},
  year={2024}
}

@inproceedings{helwig2023group,
author = {Helwig, Jacob and Zhang, Xuan and Fu, Cong and Kurtin, Jerry and Wojtowytsch, Stephan and Ji, Shuiwang},
title = {Group equivariant fourier neural operators for partial differential equations},
year = {2023},
publisher = {JMLR.org},
booktitle = {Proceedings of the 40th International Conference on Machine Learning},
articleno = {525},
numpages = {24},
location = {Honolulu, Hawaii, USA},
series = {ICML'23}
}

@article{lanthaler2024discretization,
  title={Discretization error of {Fourier} neural operators},
  author={Lanthaler, Samuel and Stuart, Andrew M and Trautner, Margaret},
  journal={arXiv preprint arXiv:2405.02221},
  year={2024}
}

@article{li2023fourier,
  title={{Fourier} neural operator with learned deformations for {PDEs} on general geometries},
  author={Li, Zongyi and Huang, Daniel Zhengyu and Liu, Burigede and Anandkumar, Anima},
  journal={Journal of Machine Learning Research},
  volume={24},
  number={388},
  pages={1--26},
  year={2023}
}

@article{li2024geometry,
  title={Geometry-informed neural operator for large-scale 3d {PDEs}},
  author={Li, Zongyi and Kovachki, Nikola and Choy, Chris and Li, Boyi and Kossaifi, Jean and Otta, Shourya and Nabian, Mohammad Amin and Stadler, Maximilian and Hundt, Christian and Azizzadenesheli, Kamyar and others},
  journal={Advances in Neural Information Processing Systems},
  volume={36},
  year={2024}
}

@article{li2020fourier,
  title={{Fourier} neural operator for parametric partial differential equations},
  author={Li, Zongyi and Kovachki, Nikola and Azizzadenesheli, Kamyar and Liu, Burigede and Bhattacharya, Kaushik and Stuart, Andrew and Anandkumar, Anima},
  journal={arXiv preprint arXiv:2010.08895},
  year={2020}
}

@article{kossaifi2023multi,
  title={Multi-Grid Tensorized {Fourier} Neural Operator for High-Resolution {PDEs}},
  author={Kossaifi, Jean and Kovachki, Nikola and Azizzadenesheli, Kamyar and Anandkumar, Anima},
  journal={arXiv preprint arXiv:2310.00120},
  year={2023}
}

@article{zhijie2022fourier,
title={{Fourier} neural operator approach to large eddy simulation of three-dimensional turbulence},
author={Li, Zhijie and Peng, Wenhui and Yuan, Zelong and Wang, Jianchun},
journal={Theoretical and Applied Mechanics Letters},
volume={12},
number={6},
pages={100389},
year={2022},
issn={2095-0349},
}

@article{bonev2023spherical,
title={Spherical {Fourier} neural operators: learning stable dynamics on the sphere},
author={Bonev, Boris and Kurth, Thorsten and Hundt, Christian and Pathak, Jaideep and Baust, Maximilian and Kashinath, Karthik and Anandkumar, Anima},
journal={Proceedings of the 40th International Conference on Machine Learning (ICML)},
volume={202},
pages={2806-2823},
year={2023},
organization={JMLR.org},
articleno={117},
}

@inproceedings{
li2020neural,
title={Neural Operator: Graph Kernel Network for Partial Differential Equations},
author={Anima Anandkumar and Kamyar Azizzadenesheli and Kaushik Bhattacharya and Nikola Kovachki and Zongyi Li and Burigede Liu and Andrew Stuart},
booktitle={ICLR 2020 Workshop on Integration of Deep Neural Models and Differential Equations},
year={2019},
url={https://openreview.net/forum?id=fg2ZFmXFO3}
}

@article{azizzadenesheli2024neural,
  title={Neural operators for accelerating scientific simulations and design},
  author={Azizzadenesheli, Kamyar and Kovachki, Nikola and Li, Zongyi and Liu-Schiaffini, Miguel and Kossaifi, Jean and Anandkumar, Anima},
  journal={Nature Reviews Physics},
  pages={1--9},
  year={2024},
  publisher={Nature Publishing Group UK London}
}

@inproceedings{lanthaler2023nonlocal,
  title={Nonlocality and Nonlinearity Implies Universality in Operator Learning},
  author={Samuel Lanthaler and Zong-Yi Li and Andrew M. Stuart},
  year={2023},
  url={https://api.semanticscholar.org/CorpusID:258331770}
}

@article{schlosshauer2019quantum,
  title={Quantum decoherence},
  author={Schlosshauer, Maximilian},
  journal={Physics Reports},
  volume={831},
  pages={1--57},
  year={2019},
  publisher={Elsevier}
}

@article{lange2024architectures,
  title={From architectures to applications: A review of neural quantum states},
  author={Lange, Hannah and Van de Walle, Anka and Abedinnia, Atiye and Bohrdt, Annabelle},
  journal={Quantum Science and Technology},
  year={2024}
}

@article{torlai2018neural,
  title={Neural-network quantum state tomography},
  author={Torlai, Giacomo and Mazzola, Guglielmo and Carrasquilla, Juan and Troyer, Matthias and Melko, Roger and Carleo, Giuseppe},
  journal={Nature physics},
  volume={14},
  number={5},
  pages={447--450},
  year={2018},
  publisher={Nature Publishing Group UK London}
}

@inproceedings{sandvik2010computational,
  title={Computational studies of quantum spin systems},
  author={Sandvik, Anders W},
  booktitle={AIP Conference Proceedings},
  volume={1297},
  number={1},
  pages={135--338},
  year={2010},
  organization={American Institute of Physics}
}

@article{preskill2018quantum,
  title={Quantum computing in the NISQ era and beyond},
  author={Preskill, John},
  journal={Quantum},
  volume={2},
  pages={79},
  year={2018},
  publisher={Verein zur F{\"o}rderung des Open Access Publizierens in den Quantenwissenschaften}
}

@incollection{feynman2018simulating,
  title={Simulating physics with computers},
  author={Feynman, Richard P},
  booktitle={Feynman and computation},
  pages={133--153},
  year={2018},
  publisher={cRc Press}
}

@article{zalka1998efficient,
  title={Efficient simulation of quantum systems by quantum computers},
  author={Zalka, Christof},
  journal={Fortschritte der Physik: Progress of Physics},
  volume={46},
  number={6-8},
  pages={877--879},
  year={1998},
  publisher={Wiley Online Library}
}

@article{lloyd1996universal,
  title={Universal quantum simulators},
  author={Lloyd, Seth},
  journal={Science},
  volume={273},
  number={5278},
  pages={1073--1078},
  year={1996},
  publisher={American Association for the Advancement of Science}
}

@article{fellous2021limitations,
  title={Limitations in quantum computing from resource constraints},
  author={Fellous-Asiani, Marco and Chai, Jing Hao and Whitney, Robert S and Auff{\`e}ves, Alexia and Ng, Hui Khoon},
  journal={PRX Quantum},
  volume={2},
  number={4},
  pages={040335},
  year={2021},
  publisher={APS}
}

@article{georgescu2014quantum,
  title={Quantum simulation},
  author={Georgescu, Iulia M and Ashhab, Sahel and Nori, Franco},
  journal={Reviews of Modern Physics},
  volume={86},
  number={1},
  pages={153--185},
  year={2014},
  publisher={APS}
}

@article{mizera2023scattering,
  title={Scattering with neural operators},
  author={Mizera, Sebastian},
  journal={Physical Review D},
  volume={108},
  number={10},
  pages={L101701},
  year={2023},
  publisher={APS}
}

@article{sachdev1999quantum,
  title={Quantum phase transitions},
  author={Sachdev, Subir},
  journal={Physics world},
  volume={12},
  number={4},
  pages={33},
  year={1999},
  publisher={IOP Publishing}
}

@article{carrasquilla2017machine,
  title={Machine learning phases of matter},
  author={Carrasquilla, Juan and Melko, Roger G},
  journal={Nature Physics},
  volume={13},
  number={5},
  pages={431--434},
  year={2017},
  publisher={Nature Publishing Group UK London}
}

@article{han2021tomography,
  title={Tomography of time-dependent quantum Hamiltonians with machine learning},
  author={Han, Chen-Di and Glaz, Bryan and Haile, Mulugeta and Lai, Ying-Cheng},
  journal={Physical Review A},
  volume={104},
  number={6},
  pages={062404},
  year={2021},
  publisher={APS}
}

@article{mohseni2022deep,
  title={Deep learning of quantum many-body dynamics via random driving},
  author={Mohseni, Naeimeh and F{\"o}sel, Thomas and Guo, Lingzhen and Navarrete-Benlloch, Carlos and Marquardt, Florian},
  journal={Quantum},
  volume={6},
  pages={714},
  year={2022},
  publisher={Verein zur F{\"o}rderung des Open Access Publizierens in den Quantenwissenschaften}
}

@article{carleo2017solving,
  title={Solving the quantum many-body problem with artificial neural networks},
  author={Carleo, Giuseppe and Troyer, Matthias},
  journal={Science},
  volume={355},
  number={6325},
  pages={602--606},
  year={2017},
  publisher={American Association for the Advancement of Science}
}

@article{daley2022practical,
  title={Practical quantum advantage in quantum simulation},
  author={Daley, Andrew J and Bloch, Immanuel and Kokail, Christian and Flannigan, Stuart and Pearson, Natalie and Troyer, Matthias and Zoller, Peter},
  journal={Nature},
  volume={607},
  number={7920},
  pages={667--676},
  year={2022},
  publisher={Nature Publishing Group UK London}
}

@article{altman2021quantum,
  title={Quantum simulators: Architectures and opportunities},
  author={Altman, Ehud and Brown, Kenneth R and Carleo, Giuseppe and Carr, Lincoln D and Demler, Eugene and Chin, Cheng and DeMarco, Brian and Economou, Sophia E and Eriksson, Mark A and Fu, Kai-Mei C and others},
  journal={PRX quantum},
  volume={2},
  number={1},
  pages={017003},
  year={2021},
  publisher={APS}
}

@article{ebadi2021quantum,
  title={Quantum phases of matter on a 256-atom programmable quantum simulator},
  author={Ebadi, Sepehr and Wang, Tout T and Levine, Harry and Keesling, Alexander and Semeghini, Giulia and Omran, Ahmed and Bluvstein, Dolev and Samajdar, Rhine and Pichler, Hannes and Ho, Wen Wei and others},
  journal={Nature},
  volume={595},
  number={7866},
  pages={227--232},
  year={2021},
  publisher={Nature Publishing Group UK London}
}

@article{tacchino2020quantum,
  title={Quantum computers as universal quantum simulators: state-of-the-art and perspectives},
  author={Tacchino, Francesco and Chiesa, Alessandro and Carretta, Stefano and Gerace, Dario},
  journal={Advanced Quantum Technologies},
  volume={3},
  number={3},
  pages={1900052},
  year={2020},
  publisher={Wiley Online Library}
}

@article{trivedi2024quantum,
  title={Quantum advantage and stability to errors in analogue quantum simulators},
  author={Trivedi, Rahul and Franco Rubio, Adrian and Cirac, J Ignacio},
  journal={Nature Communications},
  volume={15},
  number={1},
  pages={6507},
  year={2024},
  publisher={Nature Publishing Group UK London}
}

@article{king2024computational,
  title={Computational supremacy in quantum simulation},
  author={King, Andrew D and Nocera, Alberto and Rams, Marek M and Dziarmaga, Jacek and Wiersema, Roeland and Bernoudy, William and Raymond, Jack and Kaushal, Nitin and Heinsdorf, Niclas and Harris, Richard and others},
  journal={arXiv preprint arXiv:2403.00910},
  year={2024}
}

@article{white1992density,
  title={Density matrix formulation for quantum renormalization groups},
  author={White, Steven R},
  journal={Physical review letters},
  volume={69},
  number={19},
  pages={2863},
  year={1992},
  publisher={APS}
}

@article{schollwock2005density,
  title={The density-matrix renormalization group},
  author={Schollw{\"o}ck, Ulrich},
  journal={Reviews of modern physics},
  volume={77},
  number={1},
  pages={259--315},
  year={2005},
  publisher={APS}
}

@article{orus2019tensor,
  title={Tensor networks for complex quantum systems},
  author={Or{\'u}s, Rom{\'a}n},
  journal={Nature Reviews Physics},
  volume={1},
  number={9},
  pages={538--550},
  year={2019},
  publisher={Nature Publishing Group UK London}
}

@book{parkinson2010introduction,
  title={An introduction to quantum spin systems},
  author={Parkinson, John B and Farnell, Damian JJ},
  volume={816},
  year={2010},
  publisher={Springer Science \& Business Media}
}

@book{levine2009quantum,
  title={Quantum chemistry},
  author={Levine, Ira N and Busch, Daryle H and Shull, Harrison},
  volume={6},
  year={2009},
  publisher={Pearson Prentice Hall Upper Saddle River, NJ}
}

@article{williams2023unified,
  title={A unified framework for U-Net design and analysis},
  author={Williams, Christopher and Falck, Fabian and Deligiannidis, George and Holmes, Chris C and Doucet, Arnaud and Syed, Saifuddin},
  journal={Advances in Neural Information Processing Systems},
  volume={36},
  pages={27745--27782},
  year={2023}
}

@misc{liuschiaffini2024,
      title={Neural Operators with Localized Integral and Differential Kernels}, 
      author={Miguel Liu-Schiaffini and Julius Berner and Boris Bonev and Thorsten Kurth and Kamyar Azizzadenesheli and Anima Anandkumar},
      year={2024},
      eprint={2402.16845},
      archivePrefix={arXiv},
      primaryClass={cs.LG},
      url={https://arxiv.org/abs/2402.16845}, 
}

@article{Zhang2016Evaluating,
  title={Evaluating the accuracy of the dynamic mode decomposition},
  author={Hao Zhang and Scott T. M. Dawson and Clarence W. Rowley and Eric A. Deem and Louis N. Cattafesta},
  journal={Journal of Computational Dynamics},
  year={2016},
  url={https://api.semanticscholar.org/CorpusID:56094239}
}

@misc{FNOPracticalReview,
      title={Fourier Neural Operators Explained: A Practical Perspective}, 
      author={Valentin Duruisseaux and Jean Kossaifi and Anima Anandkumar},
      year={2025},
      eprint={2512.01421},
      archivePrefix={arXiv},
      primaryClass={cs.LG},
      url={https://arxiv.org/abs/2512.01421}, 
}

@inproceedings{trabelsi2018deep,
  title={Deep Complex Networks},
  author={Trabelsi, Chiheb and Bilaniuk, Olexa and Zhang, Ying and Serdyuk, Dmitriy and Subramanian, Sandeep and Santos, Joao Felipe and Mehri, Soroush and Rostamzadeh, Negar and Bengio, Yoshua and Pal, Christopher J},
  booktitle={International Conference on Learning Representations},
  year={2018}
}
\pagebreak

\appendix

\end{document}